\documentclass[showpacs,showkeys,amssymb,aps,notitlepage,twocolumn]{revtex4-1}

\usepackage{graphicx}
\usepackage{bm}

\usepackage{amsmath}

\begin{document}

\title[Ellipsoidal Geometry of Refraction in Earth's Atmosphere]{Apparent Places with an Ellipsoidal Geometry of Refraction in the Earth's Atmosphere}

\author{Richard J. Mathar}
\homepage{http://www.mpia.de/~mathar}
\email{mathar@mpia.de}
\affiliation{Max-Planck Institute of Astronomy, K\"onigstuhl 17, 69117 Heidelberg, Germany}

\date{\today}

\begin{abstract}
The displacement of star images by atmospheric refraction
observed by an Earth-bound telescope
is dominated by a familiar term proportional to the product of the tangent of the zenith angle by
the refractivity at the ground.

The manuscript focuses on the torsion of the ray path
through the atmosphere
in a model of atmospheric layers above the ellipsoidal Earth surface, induced by
the two slightly different principal curvatures along N--S and E--W pointing directions.
This breaking of the azimuthal symmetry effects apparent places at the sub-milliarcsecond scale at
optical and infrared wavelengths.

\end{abstract}

\maketitle

\section{Ray Paths}
\subsection{Tracking Snell's Law}

Tracing of rays of stellar light through the Earth atmosphere is a repeated
application of Snell's law of refraction, as the refractive index $n$ increases
from $n-1=0$ high above the atmosphere to $n-1$ of
the order of $2\times 10^{-4}$---or rather $3\times 10^{-4}$ at sea level---at the telescope site.
The angle between their direction at 
arrival at the telescope and the topocentric zenith is smaller than in a calculation
where vacuum were replacing the atmosphere \cite{RadauAnOP16,Green1985,AuerAJ119,StonePASP108,FilippenkoPASP94};
the atmosphere
is a
gradient lens \cite{MatharBA14,MarchandProgOpt11,HuiAJ572};
the ray path is curved.
Snell's law relates refractive index $n$ and incidence angle $\psi$
measured towards the normal
to the layered atmosphere:
$$
n\sin\psi = const
.
$$
The first differential of this equation is
$$
\Delta n\sin\psi +n\cos\psi \Delta \psi=0,
$$
or upon division through $\cos\psi$
\begin{equation}
\Delta n\tan\psi= -n\Delta \psi.
\label{eq.dndpsi}
\end{equation}

The gradient and the zenith vector change in direction
along the path through the atmosphere as we assume
that $n(r)$ becomes a function of the radial distance to the surface of
the Earth.
The apparent zenith angle of the ray reaching the ground, $z_0$, differs
from the zenith angle $z$ of astronomical interest (as if the action of the atmosphere was undone)
by some angle
\begin{equation}
R=z-z_0 >0.
\end{equation}
Integrating (\ref{eq.dndpsi})
its value becomes 
\cite[(4.19)]{Green1985}\cite{ThomasJHUAPL17,AuerAJ119,NenerJOSA20,NoerdlingerJPRS54,Tannousarxiv01}
\begin{equation}
R=\rho n_0\sin z_0\int_1^{n_0}dn \frac{1}{n(r^2n^2-\rho ^2n_0^2\sin^2z_0)^{1/2}},
\label{eq.RofnInt}
\end{equation}
obtained in radian.

This integral over the refractive index starts above the atmosphere
where $n=1$, and ends at the telescope position where $n=n_0$ (refractive index on the ground).
$r\ge \rho$ is the distance to the Earth center.

The binomial expansion of the integrand of $R$ in powers of
$\sin z_0$ is \cite[(3.6.12)]{AS}
\begin{eqnarray}
\frac{1}{\sqrt{a^2-b^2 \sin^2 z_0}}
\approx
&&
\frac{1}{a}\Big[1+\frac{b^2}{2a^2}\sin^2 z_0 +\frac{3b^4}{8a^4}\sin^4 z_0
\nonumber \\
&&
+\frac{5b^6}{16a^6}\sin^6 z_0+\cdots\Big]
\end{eqnarray}
and leads to
\begin{eqnarray}
R\approx n_0 \sin z_0
&&\Big[I_0+\frac{1}{2}I_1\sin^2 z_0+\frac{3}{8}I_2\sin^4 z_0
\nonumber \\
&&+\frac{5}{16}I_3\sin^6z_0+\cdots\Big]
\label{eq.Rofsinz}
\end{eqnarray}
in terms of dimensionless integrals
\begin{equation}
I_k\equiv \int_1^{n_0}\frac{\rho (\rho n_0)^{2k}}{rn^2(rn)^{2k}}dn
.
\label{eq.Ik}
\end{equation}

The coefficients $I_k$ of that expansion turn out to change weakly as
a function of the index, only by a few percent.
So the convergence of the series is usually accelerated by switching
from the sines to the tangents \cite[(3.6.12),(4.3.45)]{AS},
\begin{eqnarray}
\sin z_0=\frac{\tan z_0}{\sqrt{1+\tan^2z_0}}
\approx
&&
\tan z_0\Big[1-\frac{1}{2}\tan^2z_0+\frac{3}{8}\tan^4z_0
\nonumber \\
&&
-\frac{5}{16}\tan^6z_0+-\cdots\Big],
\end{eqnarray}
which recombines terms in (\ref{eq.Rofsinz}) as \cite{StonePASP108,MatharBA14}
\begin{eqnarray}
R\approx && n_0 \tan z_0
\Big[I_0
-\frac{1}{2}(I_0-I_1)\tan^2 z_0
\nonumber \\
&& +\frac{3}{8}(I_0-2I_1+I_2)\tan^4 z_0
\nonumber \\
&&
-\frac{5}{16}(I_0-3I_1+3I_2-I_3)\tan^6 z_0+\cdots
\nonumber \\
&& +(-)^l\frac{(2l-1)!!}{(2l)!!}\sum_{k=0}^l
\binom{l}{k}
(-)^kI_k\tan^{2l} z_0+\cdots
\Big].
\label{eq.Roftanz}
\end{eqnarray}

Starting from these well-known arguments,
the following chapters quantify how this distortion $R$ of the sky
map in the observer's spherical projection is not only a function of 
the zenith angle but also of the azimuth, because the path integrals
$I$ depend on the curvature of the atmospheric layers imprinted by
the Earth curvature $1/\rho$ and its weak variations implied by an ellipsoidal
model of the Earth surface.

\subsection{Refractive Index of Altitude}

Accurate modeling of $n(r)$ is not within the scope of this treatise here.
The algebra that follows uses an exponential model of the air susceptibility $\chi$
as a function of altitude
with a scale height $K$,
\begin{equation}
n = 1+\frac{\chi_0}{2} e^{-(r-\rho)/K}, \quad
n_0=1+\frac{\chi_0}{2}.
\label{eq.chiofr}
\end{equation}
The parameters were set to
\begin{equation}
\chi_0=4\times 10^{-4},\quad \rho=6380\, \mathrm{km},\quad K=9.6\, \mathrm{km},
\label{eq.K}
\end{equation}
representing prototypical values in the near infrared at $h=2600$ m
above sea level.

We substitute a dimensionless air mass
$H\equiv (r-\rho)/K$
and its differential
$dn/dH = -\frac{\chi_0}{2}e^{-H}$
into (\ref{eq.Ik}) and (\ref{eq.chiofr}), so we aim to calculate
in order $k=0,1,2,\ldots$
\begin{equation}
I_k
=
\frac{\chi_0}{2}n_0^{2k}\int_0^ \infty\frac{e^{-H}}{(1+HK/\rho)^{2k+1}(1+\frac{\chi_0}{2}e^{-H})^{2k+2}} dH
.
\nonumber
\end{equation}
The denominator is expanded in a power series of the small
unitless parameter $\hat K\equiv K/\rho\approx 0.0015$:
\begin{equation}
\frac{1}{(1+H\hat K)^t}
=
\sum_{m\ge 0}\binom{t+m-1}{m}(-H\hat K)^m
.
\end{equation}
Rewriting the altitude integral as in
\ref{sec.AppU} decomposes the path integrals in orders $m$ of the curvature:
\begin{eqnarray}
I_k
&=&
\frac{\chi_0}{2}n_0^{2k} \sum_{m\ge 0}
\binom{2k+m}{m}
(-\hat K)^m
\int_0^ \infty\!\!\! H^m \frac{e^{-H}dH}{(1+\frac{\chi_0}{2}e^{-H})^{2k+2}}
\nonumber
\\
&=&
\frac{\chi_0}{2}n_0^{2k} \sum_{m\ge 0}
\binom{2k+m}{m}
(-\hat K)^m
U_{m,2k+2}(\chi_0/2)
\nonumber
\\
&\equiv &
\sum_{m\ge 0} I_{k,m}
.
\label{eq.IofU}
\end{eqnarray}
Thanks to the simplicity of the exponential refractivity
model (\ref{eq.chiofr}),
this strategy leads to a manageable numerical
implementation \emph{and} conveniently
encapsulates the small effect of variations of the Earth
radius $\rho$ in the leading terms of small $m$.
In view of the series (\ref{eq.Upowalpha}),
$I_k$ is actually represented as a bivariate power series
of $\hat K$ (curvature) and $\chi_0/2$ (refractivity).

\section{Geodetic Earth Coordinates}

\subsection{Oblate Ellipsoid Coordinates}
The Cartesian coordinates of a position $\bf r$ are related to the
geodetic latitude $\phi$, geodetic longitude $\lambda$ and height $h$ above
an oblate ellipsoid as \cite{JonesJG76}
\begin{equation}
\bf{r} =\left(
\begin{array}{c}
(N+h)\cos\phi\cos\lambda\\
(N+h)\cos\phi\sin\lambda\\
\left[N(1-e^2)+h\right]\sin\phi
\end{array}
\right),
\end{equation}
where
\begin{equation}
N(\phi)\equiv \frac{\rho_e}{\sqrt{1-e^2\sin^2\phi}}
\label{eq.Ndef}
\end{equation}
is the distance to the Earth axis along the surface normal. $\rho_e\approx 6378$ km
and $e\approx 0.0818$ are equatorial radius
and eccentricity \cite{McCarthyIERS32}.
A polar radius is defined as $\rho_p=\rho_e/\sqrt{1-e^2}$ \cite{MoritzBullG54}.

\subsection{Azimuth-Dependent Curvatures} \label{sec.curv}
The Fundamental Parameters of the first form of the surface of constant altitude
$h$ for prolate ellipsoids are \cite{MatharArxiv0711}
\begin{eqnarray}
\bar E&=& (N+h)^2\cos^2\phi ;
\\
\bar F&=& 0 ;
\\
\bar G&=& (M+h)^2 .
\end{eqnarray}
The Fundamental Parameters of the second form of the surface are \cite{HarrisOPO26}
\begin{eqnarray}
\bar L&=& -(N+h)\cos^2\phi ;
\\
\bar M&=& 0 ;
\\
\bar N&=& -(M+h) ,
\end{eqnarray}
where
\begin{equation}
M(\phi)\equiv N(\phi)\frac{1-e^2}{1-e^2\sin^2\phi}.
\label{eq.Mdef}
\end{equation}
The six Fundamental Parameters are hatted with an overbar
to set $\bar N$ and $\bar M$ apart from the
distances $N$ and $M$ in (\ref{eq.Ndef}) and (\ref{eq.Mdef}).
The two principal curvatures $\kappa_{1,2}(\phi)$ are the
roots of the quadratic equation
\begin{equation}
(\bar E\bar G-\bar F^2)\kappa^2
-(\bar E\bar N-2\bar F\bar M+\bar G\bar L)\kappa
+(\bar L\bar N-\bar M^2)=0.
\end{equation}
After insertion of the six parameters, the quadratic equation reads
\begin{equation}
(N+h)(M+h) \kappa^2
+(M+N+2h)\kappa
+1=0.
\label{eq.kappa}
\end{equation}

\begin{figure}[h]
\includegraphics[width=0.45\textwidth]{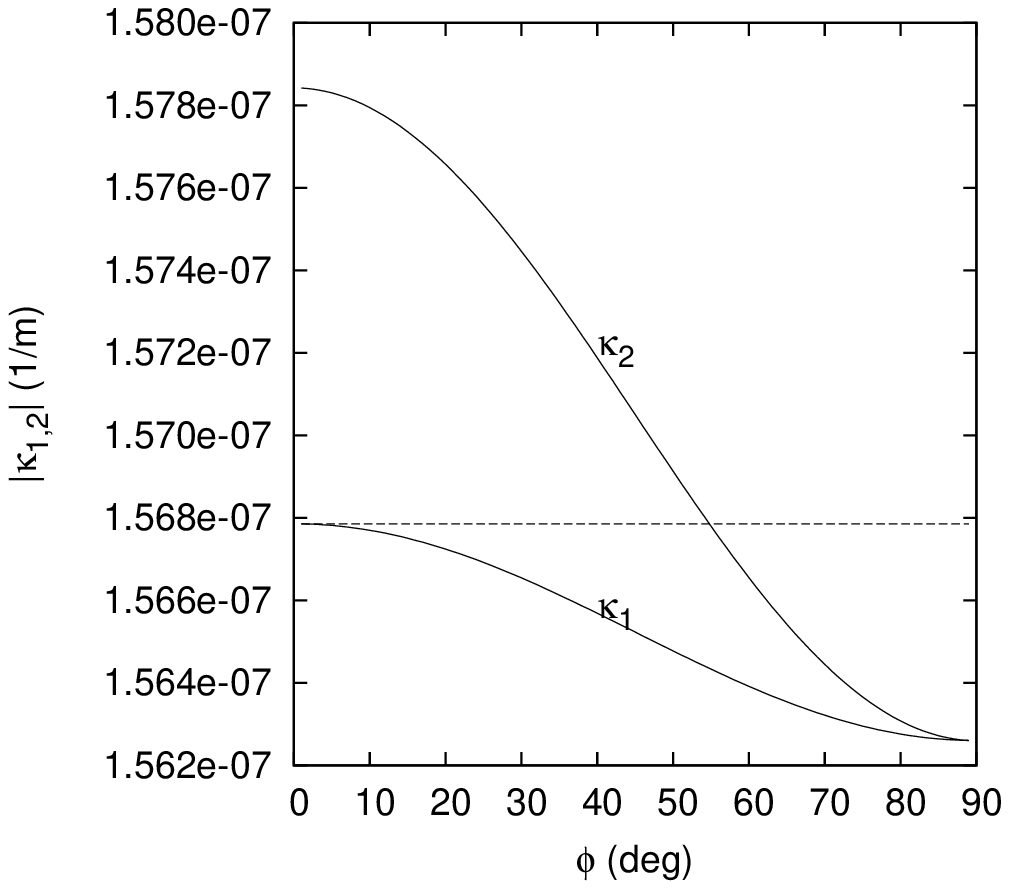}
\caption{
\label{fig:curv}
The absolute values of the two principal curvatures $\kappa$ (\ref{eq.kappas}) at $h=0$
as a function of geodetic latitude $\phi$ from the equator to the poles.
The dashed line indicates the value $|\kappa_1(\phi=h=0)|=1/\rho_e$.
}
\end{figure}
Its two solutions
\begin{equation}
\kappa_1=-1/(N+h),\quad \kappa_2=-1/(M+h)
\label{eq.kappas}
\end{equation}
are shown in Figure \ref{fig:curv}.
They unite at the poles:
\begin{equation}
\kappa_{1,2}(\phi=\pm \pi/2)
=-\frac{1}{\rho_p+h}.
\label{eq.kappPol}
\end{equation}

We call $|\kappa|$ the curvature, but
coherent with the standard nomenclature for closed shell surfaces,
the negative sign is maintained in the values.

Euler's formula for the curvature as a function of azimuth angle $A$ in 
the topocentric coordinate system is
\begin{equation}
\kappa= \kappa_1\sin^2 A+\kappa_2\cos^2 A
=\frac{\kappa_1+\kappa_2}{2}+\frac{\kappa_2-\kappa_1}{2}\cos(2A).
\label{eq.Eul}
\end{equation}
The argument $2A$ of the cosine illustrates that it does not matter whether azimuths are measured
relative to South or North directions  in the observer's topocentric tangent plane.
The expansion of the amplitude term in a power series of the cosine
of the geodetic latitude is to fourth order
\begin{eqnarray}
\frac{\kappa_2-\kappa_1}{2}
& \approx &
-\frac{\rho_p}{2(1-e^2)(\rho_p+h)^2}(e\cos\phi)^2
\nonumber
\\
&&
-\frac{\rho_p(\rho_p-3h)}
{4(1-e^2)^2(\rho_p+h)^3}(e\cos\phi)^4
,
\label{eq.kampl}
\end{eqnarray}
where (\ref{eq.Ndef}) and (\ref{eq.Mdef}) have been inserted into (\ref{eq.kappas}).

\section{Synopsis}
\subsection{Rules of Thumb}

The previous two sections considered (i) the zenith dependence of the apparent zenith angle
calculated from a functional of the refractive index and
(ii) the weak dependence of the Earth surface curvature
on the pointing azimuth. The central theme of this
letter is to combine both aspects by imposing the second aspect
on the altitude model of the atmospheric thickness in the first aspect.

The sensitivity of the integrals $I_k$ and eventually of the apparent zenith
angle on the Earth radius and therefore on the curvature $\kappa$ (represented
by $\hat K$) is easily estimated as follows:
The derivatives of (\ref{eq.IofU}) are
$$
\frac{\partial I_k}{\partial \kappa}
=
\frac{\chi_0}{2}n_0^{2k} K\sum_{m\ge 1} m\binom{2k+m}{m}(-\hat K)^{m-1}U_{m,2k+2}(\chi_0/2)
,
$$
and the dominating order is at $m=1$
\begin{equation}
\frac{\partial I_k}{\partial \kappa}
\approx
\frac{\chi_0}{2}n_0^{2k} (2k+1) K U_{1,2k+2}(\chi_0/2)
\label{eq.Ikdelta}
.
\end{equation}
All $U_{m,s}$ are very close to $m!$ [see (\ref{eq.Upowalpha})]
and $n_0$ is approximately one, so the contribution from $I_{k=0}$
to the differential is approximately
$
\frac{\chi_0}{2}K$.
The change $\Delta I_0\approx (\partial I_0/\partial \kappa)\Delta \kappa$
induced by an azimuthal curvature variation is
$\Delta I_0\approx 1.9\times 10^{-9}$ rad $\approx 0.4$ milli-arcseconds (in the order $\tan z_0$),
taking parameters from (\ref{eq.K}) and $\Delta \kappa\le 0.01\times 10^{-7}$ 1/m
from Figure \ref{fig:curv}.
$\Delta \kappa$ is limited by two times the maximum amplitude in 
\ref{eq.kampl}, approximately $e^2/\rho_p$.
The factor $2k+1$ in (\ref{eq.Ikdelta}) means that the effect on
$I_{k=1}$ is approximately three times of this, so the effect on the combined $-(I_0-I_1)/2$
is approximately the same as the effect on $I_0$ alone; in the order $\tan^3z_0$ in (\ref{eq.Roftanz})
this accounts for another 0.4 milli-arcseconds at $z_0=45^\circ$ or 0.7 milli-arcseconds at $z_0=60^\circ$.
The equivalent reasoning with $I_2$, $I_3$ and higher orders indicates that the contributions
from the fifth and higher orders of $\tan z_0$ are negligible, because
the factors $2k+1$ are annihilated in the alternating sign sums over 
the rows over the Pascal triangle times $I_k$ in (\ref{eq.Roftanz}):
$\sum_{k=0}^l\binom{l}{k}(-1)^k(2k+1)=0$ if $l>1$.
In total, the
undulation of $R$
along a full sweep of the azimuth is roughly limited to 1.1 milli-arcseconds for zenith angles
below $60^\circ$.

\subsection{Numerics}
Plugging (\ref{eq.IofU}) into (\ref{eq.Roftanz}) represents $R$ by a double sum over the orders $l$ and $m$.
Direct numerical computation outlined in Appendix \ref{app.prog}
creates Table \ref{tab.RofI}, which summarizes the contribution of tangent order 
$2l+1$ to $R$ refined along the curvature orders $m$.

\begin{table*}
\caption{
\label{tab.RofI}
The value of $\frac{(2l-1)!!}{(2l)!!}\sum_k \binom{l}{k} (-)^{k+l}I_{k,m}(\alpha)$ at 
$\alpha=2\times 10^{-4}$ for rows $l=0\ldots 4$
and columns $m$ contributing to (\ref{eq.Roftanz})
through the curvature order $O({\hat K}^m)$.
}
\begin{ruledtabular}
\begin{tabular}{r|llllll}
$l\backslash m $ & 0 & 1 & 2 & 3 \\
\hline
0  & $1.99960\times 10^{-4}$  & $-3.00910\times 10^{-7}$  & $9.05606\times 10^{-10}$  & $-4.08810\times 10^{-12}$ \\
1  & $1.99973\times 10^{-8}$  & $-3.01046\times 10^{-7}$  & $2.26497\times 10^{-9}$  & $-1.84041\times 10^{-11}$ \\
2  & $3.99980\times 10^{-12}$  & $-1.35474\times 10^{-10}$  & $1.36055\times 10^{-9}$  & $-2.45574\times 10^{-11}$ \\
3  & $1.00004\times 10^{-15}$  & $-5.51972\times 10^{-14}$  & $1.19009\times 10^{-12}$  & $-1.02548\times 10^{-11}$ \\
4  & $2.80037\times 10^{-19}$  & $-2.19553\times 10^{-17}$  & $7.49284\times 10^{-16}$  & $-1.34458\times 10^{-14}$ \\
\end{tabular}
\end{ruledtabular}
\end{table*}

The table demonstrates that the two leading coefficients from the
first curvature order $m=1$ are already larger than the coefficient
of the tangent order $l=1$ derived cojointly with a  planar surface at $m=0$.
In practise this means that one ought---at least roughly---consider the
curvature to first order \emph{if} one aims to look at the cubic
or higher orders of the tangent expansion (\ref{eq.Roftanz}).
The combined associated zenith angle correction of $\approx 2\times 3\times 10^{-7}$ radians
is approximately 0.12 arcseconds at $z=45^\circ$,
and for example
noticeable in comparison to the pointing accuracy from encoder errors
measured with the telescopes of the Very Large Telescope \cite{WallanderSPIE4004,KoehlerSPIE6268}.

The column $m=1$ supports the assessment in the previous section that the contribution
from $I_{0,m}$ and the combined $I_{0,m}-I_{1,m}$
in Table \ref{tab.RofI} are almost the same, $3\times 10^{-7}$,
and much larger than the entries
further down in that column.

The sweeped variation estimated in the previous section
is merely the up to 0.6 percent effect of variation
in $\kappa$ illustrated in Figure \ref{fig:curv} multiplied
as a linear perturbation by the numbers of column $m=1$.

\subsection{Underlying Principles}
The difference $z-z_0$ is the familiar change in zenith angle. The
dependence on the azimuth $A$ measures the ray torsion \cite{HubbardIcar27,SyndergaardJATS60}, which vanishes in the
limit of zero eccentricity.

The presence of torsion---meaning rays do not stay in a common plane of incidence
 defined above the atmosphere---is a
side
effect of the different curvatures of the atmospheric layers
along the N--S and E--W directions set up by this atmospheric model.
The (absolute value of the) curvature is larger for pointing into N--S than for pointing into E--W directions. Larger
curvature implies smaller air mass,
because the atmosphere nose-dives quicker to the horizon,
implies smaller angles of incidence relative to
the normals to the air layers,
and implies smaller optical path length.
The different signs of the entries in the left columns in Table \ref{tab.RofI} witness that the
curvature term $(m=1)$ slightly weakens the bulk air mass term $(m=0)$,
compliant with this argument of signs.

Since Snell's law is a result of the
eikonal minimization (Fermat's principle) which lets light select
minimum optical path length,
the net effect is to steer light closer to the N if it has a northern component,
and closer to S if it has southern component.

The effect vanishes if the telescope is near the geodetic poles, $\phi\to \pm 90^\circ$,
because
the two curvatures equalize there (equation (\ref{eq.kappPol})).

\section{Summary} 
Ray paths
through the atmosphere
in the geometric optics approximation
change their direction
in zenith angle \emph{and} azimuth if the
refractive indices in the atmosphere are
smoothly bent along
the prolate ellipsoid of the Earth.
The scale shift as the azimuth angle changes is less than one milli-arcsecond
for pointing less than $60^\circ$ away from the zenith at typical optical
wavelengths,
and therefore much smaller than the standard refractivity effect
which changes apparent positions on a scale of 60 arcseconds
under the same conditions.

\appendix
\section{Auxiliary Air Mass Integral}\label{sec.AppU}
The integral over the exponential index of refraction in (\ref{eq.IofU}) is
\begin{equation}
U_{m,s}(\alpha)\equiv \int_0^\infty H^m \frac{e^{-H}}{(1+\alpha e^{-H})^s}dH
=
\int_0^1 \frac{(-\log t)^m}{(1+\alpha t)^s}dt
\label{eq.Ulogm}
\end{equation}
in terms
of an optical thickness parameter $t\equiv e^{-H}$.

On the lattice of the two integer parameters $m,s\ge 0$
some elementary cases emerge \cite[2.117.2,2.711]{GR}:
\begin{equation}
U_{0,1}(\alpha)=
\int_0^1 \frac{1}{1+\alpha t}dt
=
\frac{1}{\alpha}\ln (1+\alpha),
\end{equation}
\begin{equation}
U_{0,s}(\alpha)
=
\frac{1}{\alpha(s-1)} [1-\frac{1}{(1+\alpha)^{s-1}}],\quad s>1,
\label{eq.U0s}
\end{equation}
\begin{equation}
U_{m,0}(\alpha) = m!,
\end{equation}
which includes the case $0!=1$.

With the aid of \cite[2.711]{GR}
\begin{equation}
\int (-\log t)^m dt
=
t (-\log t)^m + t \sum_{l=1}^m \frac{m!}{(m-l)!}(-\log t)^{m-l}
\end{equation}
and partial integration of (\ref{eq.Ulogm}) we arrive for $m>0$ at:
\begin{eqnarray}
U_{m,s}(\alpha)
=
&&
m! \frac{1}{(1+\alpha)^s}
+
s
\Big\{
U_{m,s}-U_{m,s+1}
\nonumber \\ &&
+\sum_{l=1}^m \frac{m!}{(m-l)!}
[U_{m-l,s}-U_{m-l,s+1}]
\Big\}.
\end{eqnarray}
This offers to bootstrap values at $m>0$ or $s>0$ via the recurrence
\begin{eqnarray}
sU_{m,s+1}=
&&
m! \frac{1}{(1+\alpha)^s}
+
(s-1)
U_{m,s}
\nonumber \\
&&
+s\sum_{l=1}^m \frac{m!}{(m-l)!}
[U_{m-l,s}-U_{m-l,s+1}]
.
\end{eqnarray}

\begin{table}
\caption{
The value of $\binom{2k+m}{m}U_{m,s}(\alpha)$ at $\alpha=2\times 10^{-4}$ for rows $m=0\ldots 5$
and columns $s=2k+2=2,4 $ and $6$.
}
\begin{ruledtabular}
\begin{tabular}{r|rrrrrr}
$m\backslash s$ & 2 & 4 & 6 \\
\hline
0 & 0.9998000400 & 0.9996001333 & 0.9994002799 \\
1 & 0.9999000133 & 2.9994001333 & 4.9985004665 \\
2 & 1.9999000089 & 11.9988001777 & 29.9955009331 \\
3 & 5.9998500089 & 59.9970002963 & 209.9842521774 \\
4 & 23.9997000119 & 359.9910005925 & 1679.9370058067 \\
5 & 119.9992500198 & 2519.9685013826 & 15119.7165174206 \\
\end{tabular}
\end{ruledtabular}
\end{table}

The power series of (\ref{eq.Ulogm}) is \cite[2.722]{GR}
\begin{equation}
U_{m,s}
=
m!
\sum_{l\ge 0}
(-\alpha)^l
\binom{s+l-1}{l}
\frac{1}{(l+1)^{m+1}}
.
\label{eq.Upowalpha}
\end{equation}

Remark: (\ref{eq.U0s}) establishes the special value $U_{0,2}(\alpha)=1/(1+\alpha)=1/n_0$.
The dominant contribution in (\ref{eq.IofU}) stems from $m=0$,
so $I_0\approx \frac{\chi_0}{2}U_{0,2}(\chi_0/2)=\chi_0/(2n_0)$.
In consequence the dominant contribution
from that leading term
to (\ref{eq.Roftanz})
is $R\approx n_0I_0\tan z_0\approx(n_0-1)\tan z_0$.
This is the standard formula of the literature in the limit of
a flat Earth surface
\cite{Green1985,FilippenkoPASP94}.

\section{Programmer's Guide}\label{app.prog}
The steps of a numerical computation of the ``transverse'' atmospheric
dispersion $R$ are:
\begin{enumerate}
\item
Fix the set of Earth coordinates for the prolate ellipsoid:
equatorial radius $\rho_e$, and eccentricity $e$;
\item
Fix the observer's coordinates in that reference system: altitude $h$
above the ellipsoid, and geodetic latitude $\phi$;
\item
Compute the distances $M$ and $N$ from (\ref{eq.Ndef}) and (\ref{eq.Mdef});
\item
Fix the
pointing direction
in the observer's topocentric coordinate
system: zenith angle $z_0$, and azimuth $A$;
\item Compute the value $\kappa$ of the curvature for that pointing
direction with (\ref{eq.Eul});
\item Fix the model parameters for the air refractivity:  the scale
height $K$, and the index of refraction $n_0$ on the ground;
\item
Compute $\alpha=n_0-1$ at the wavelength of interest and a table of $U_{m,2k+2}(\alpha)$ 
for small $m,k\ge 0$
as outlined in
\ref{sec.AppU};
\item
Compute the parameter $\hat K=K|\kappa|$ and derive the table
of $I_k$ from (\ref{eq.IofU});
\item
Insert these $I_k$ into (\ref{eq.Roftanz}).
\end{enumerate}

\bibliographystyle{apsrmp4-1}
\bibliography{all}

\end{document}